\RequirePackage{amsmath}
\documentclass{llncs}

\usepackage{amsmath}
\usepackage{amssymb}
\usepackage{array}
\usepackage{booktabs}
\usepackage{color}
\usepackage{float}
\usepackage{graphicx}
\usepackage{ifpdf}
\usepackage[utf8]{inputenc}
\usepackage{keyval}  
\usepackage{listings}   
\usepackage{moresize}
\usepackage{multirow}
\usepackage[numbers,sort&compress,square]{natbib}
\usepackage{paralist}    
\usepackage{rotating}
\usepackage{soul}
\usepackage{srcltx}
\usepackage{url}
\usepackage{xcolor}
\usepackage{xspace}
\usepackage{wrapfig}
\usepackage{subfig}
\usepackage[export]{adjustbox}

\definecolor{listinggray}{gray}{0.95}
\definecolor{darkgray}{gray}{0.7}
\definecolor{commentgreen}{rgb}{0, 0.4, 0}
\definecolor{darkblue}{rgb}{0, 0, 0.6}
\definecolor{purple}{rgb}{0.6, 0, 0.6}
\definecolor{middleblue}{rgb}{0, 0, 0.75}
\definecolor{darkred}{rgb}{0.4, 0, 0}
\definecolor{brown}{rgb}{0.5, 0.5, 0}
\definecolor{dkgreen}{rgb}{0,0.5,0}
\definecolor{orange}{rgb}{1,.5,0}
\definecolor{dandelion}{cmyk}{0,0.29,0.84,0}

\usepackage[normalem]{ulem}
\makeatletter
\def\cyanuwave{\bgroup \markoverwith{\lower3.5\p@\hbox{\sixly \textcolor{cyan}{\char58}}}\ULon}
\def\reduwave{\bgroup \markoverwith{\lower3.5\p@\hbox{\sixly \textcolor{red}{\char58}}}\ULon}
\def\blueuwave{\bgroup \markoverwith{\lower3.5\p@\hbox{\sixly \textcolor{blue}{\char58}}}\ULon}
\font\sixly=lasy6 
\makeatother

\newif\ifdraft{}

\ifdraft{}
  \newcommand{\amnote}[1]{ \textcolor{blue} { ***andrem: #1 }}
  \newcommand{\jhanote}[1]{ {\textcolor{red} { ***shantenu: #1 }}}
  \newcommand{\mtnote}[1]{ {\textcolor{orange} { ***matteo: #1 }}}
  \newcommand{\mmnote}[1]{ {\textcolor{violet} { ***manuel: #1 }}}
\else
  \newcommand{\amnote}[1]{}
  \newcommand{\jhanote}[1]{}
  \newcommand{\mtnote}[1]{}
  \newcommand{\mmnote}[1]{}
\fi

\newcommand{\B}[1]{\textbf{#1}\xspace}

\newcommand{\I}[1]{\textit{#1}\xspace}
\newcommand{\T}[1]{\texttt{#1}\xspace}

\newcommand{\UPP}{\vspace*{-2.0em}}
\newcommand{\UP}{\vspace*{-1.0em}}
\newcommand{\up}{\vspace*{-0.5em}}

\newcommand{\mr}[1]{\multirow{2}{*}{#1}}%

\lstdefinestyle{myListing}{
  frame=single,   
  backgroundcolor=\color{listinggray},  
  language=C,       
  basicstyle=\ttfamily \footnotesize,
  breakautoindent=true,
  breaklines=true
  tabsize=2,
  captionpos=b,  
  aboveskip=0em,
  belowskip=-2em,
}      

\lstdefinestyle{myPythonListing}{
  frame=single,   
  backgroundcolor=\color{listinggray},  
  language=Python,       
  basicstyle=\ttfamily \footnotesize,
  breakautoindent=true,
  breaklines=true
  tabsize=2,
  captionpos=b,  
}

\ifpdf{}
  \DeclareGraphicsExtensions{.pdf, .jpg, .tif}
\else
  \DeclareGraphicsExtensions{.eps, .jpg, .ps}
\fi

\title{Design and Performance Characterization of RADICAL-Pilot on Titan}

\author{Andre Merzky$^{1}$ \and  Matteo Turilli$^{1}$ \and Manuel Maldonado$^{1}$ \and Shantenu Jha$^{1,2}$}

\institute{$^{1}$RADICAL Laboratory, Electric and Computer Engineering, \\
  Rutgers University, New Brunswick, NJ, USA \\ $^{2}$ Brookhaven National Laboratory, Upton, New York, USA}

\begin{document}
\maketitle

\begin{abstract}
Many extreme scale scientific applications have workloads comprised of a
large number of individual high-performance tasks. The Pilot abstraction
decouples workload specification, resource management, and task execution via
job placeholders and late-binding. As such, suitable implementations of the
Pilot abstraction can support the collective execution of large number of
tasks on supercomputers. We introduce RADICAL-Pilot (RP) as a portable,
modular and extensible Python-based Pilot system. We describe RP's design,
architecture and implementation. We characterize its performance and show its
ability to scalably execute workloads comprised of thousands of MPI tasks on
Titan---a DOE leadership class facility. Specifically, we investigate RP's
weak (strong) scaling properties up to 131K (65K) cores and 4096 (16384) 32
core tasks. RADICAL-Pilot can be used stand-alone, as well as integrated with
other tools as a runtime system.
\end{abstract}

\section{Introduction}\label{sec:intro}

Supercomputers have been designed to support applications comprised of mostly
monolithic, single-job workloads. Many existing scientific applications have
workloads comprised of multiple heterogeneous tasks. A computational task is
a generalized term for a stand-alone process that has well defined input,
output, termination criteria, and dedicated resources. A task can be used to
represents an independent simulation or data processing analysis, running on
one or more nodes of a high-performance computing machine, that may or not
use MPI\@. The number of such applications account for an increasing fraction
of high-performance computing utilization~\cite{nersc-analysis-14,multiple1}.

Further, supercomputers are typically operated to maximize overall system
utilization, which entails static resource partitioning across jobs and
users. Thus, there is a tension between the resource requirements of
\textit{non-traditional} applications comprised of many smaller tasks, and
the capabilities of the traditional HPC system software as well as their
usage policies. Whether these applications directly access supercomputers, or
use workflow systems, they could benefit from better execution and resource
management on HPC resources~\cite{better-resource}. The increasing number of
systems (\S\ref{sec:related}) that address this problem is testimony to its
importance.

Pilot systems~\cite{review_pilotreview} address two apparently contradictory
requirements: accessing HPC resources via their centralized schedulers while
letting applications independently schedule tasks on the acquired portion of
resources. By implementing multi-level scheduling and late-binding, Pilot
systems lower task scheduling overhead, enable higher task execution
throughput, and allow greater control over the resources acquired to execute
workloads. As such, Pilot systems provide a promising starting point to ease
the tension between the resource requirements of workloads comprised of
multiple tasks and the the capabilities of the traditional HPC system
software.

We introduce RADICAL-Pilot (RP), a Pilot system that implements the pilot
paradigm as outlined in Ref.~\cite{review_pilotreview}. RP is implemented in
Python and provides a well defined API and usage modes. Although RP is
vehicle for research in scalable computing, it also supports production grade
science. Currently, it is being used by applications drawn from diverse
domains, ranging from earth sciences and biomolecular sciences to high-energy
physics. RP directly supports their use of supercomputers or it can be used
as a runtime system by third party workflow or workload management 
systems~\cite{bb_2016,entk,repex2016,power-of-many17,dakka2017htbac,htchpc2017converging,turilli2016analysis}.

RP is not a workflow system and does not provide workflow or workload
management capabilities itself. In 2017, RP was used to support more than
100M core-hours on US DOE, NSF resources (Blue Waters and XSEDE), and
European supercomputers (Archer and SuperMUC). In 2018, RP is the core
runtime system for three DOE INCITE and an NSF PRAC award consuming an
estimated lower bound of 250M core hours on several HPC machines, including
Titan, Blue Waters (NCSA), and XSEDE resources (e.g., Stampede).

A primary contribution of this paper is the investigation and fine-grained
characterization of the performance and scaling of RP to execute workloads
comprised of thousands of MPI tasks on Titan, managed at the Oak Ridge
Leadership Computing Facility. Consequently, we are able to localize the
overheads to specific components (sub-systems) of RP and thereby optimize its
performance. Specifically, we investigate RP's weak (strong) scaling
properties up to 131K (65K) cores and 4096 (16384) 32 core tasks.

Although RP works on multiple platforms, we focus our experiments on
Titan---a DOE leadership class facility, as it currently offers the highest
degree of concurrent execution in the USA for open/academic research
(approximately 300K CPU cores). On Titan, we optimized RP to overcome
existing bottlenecks, so that both the performance and scalability of RP is
determined by system software limits. Specifically, we show that the launch
rate of tasks is dominated by an overhead arising from the use of OpenMPI
Runtime Environment (ORTE) of OpenMPI\@. We also observe failures at the ORTE
layer when utilizing more than 131K cores.

In Section~\ref{sec:related}, we discuss existing pilot-systems and highlight
the distinctive capabilities of RP\@. Section~\ref{sec:arch} discusses the
design and architecture of RP\@. Section~\ref{sec:exp} describes the core
experiments and results of the paper.

\section{Related Work}\label{sec:related}

Ref.~\cite{review_pilotreview} established a distinction between the Pilot
paradigm and the Pilot abstraction. The Pilot paradigm refers to the
execution of a workload via multi-entity and multi-stage scheduling on
resource placeholders. The Pilot abstraction is associated with logical
components and functionalities of software systems. By design, the Pilot
paradigm enables the handling of multiple types of tasks, regardless of their
size and duration. There is an apparent gap in the space of Pilot system
implementations, as they have not exploited the full generality of the Pilot
paradigm.

Ref.~\cite{review_pilotreview} presents around twenty systems that have
implemented core Pilot abstraction functionality (i.e., Pilot systems) since
1995. Most of these systems are tailored to specific workloads, resources,
interfaces, or development models. Most Pilot systems have been implemented
to optimize the throughput of single-core (or single-node), short-lived,
uncoupled tasks~\cite{review_pilotreview}. For example, HTCondor with Glidein
on OSG~\cite{pordes2007open} is one of the most widely used Pilot systems but
serves mostly single core workloads; or the Pilot systems developed for the
LHC communities which execute millions of jobs a
week~\cite{maeno2014evolution} specialize on supporting LHC workloads and, in
most cases, specific resources like those of WLCG\@.

Away from distributed high-throughput pilots, the light-weight execution
framework called Falkon represents an early stand-alone Pilot system for HPC
environments~\cite{raicu2007falkon}. It implements concurrency at multiple
levels (e.g., dispatching, scheduling, etc.) and was optimized for single
core applications. Falkon's design allows it to achieve great performance by
leaving out non-essential features like, for example, the support for
multi-node tasks such as MPI~\cite{1362680}.

The Pilot abstraction has also been implemented by various workflow
management systems: for example, Pegasus~\cite{deelman2015} uses Glidein via
providers like Corral~\cite{deelman2005pegasus};
Makeflow~\cite{albrecht2012makeflow} and FireWorks~\cite{jain2015fireworks}
enable users to manually start workers on HPC resources via master/worker
tools called Work Queue~\cite{bui2011work} or
LaunchPad~\cite{jain2015fireworks}; and Swift~\cite{wilde2011swift} can use
Falkon~\cite{raicu2007falkon} and Coasters~\cite{hategan2011coasters} as
Pilot systems. In most cases, the logical components that implement the Pilot
abstraction are not stand-alone, nor is the functionality isolated in these
workflow management systems. This can limit portability and re-usability.

JETS~\cite{Wozniak:2013Jets} is a middleware component that provides high
performance support for many-parallel-task computing. It is designed for
running short-duration MPI tasks, down to the order of seconds. Workloads can
be codified in the Swift scripting language, and can be executed either by
stand-alone workers (via MPICH2~\cite{mpich2}) or by pilots managed by
Coasters. Given JETS focus on scalability of tasks with duration in the range
of seconds, resource utilization is reduced at larger number of nodes or
processes per node due to the increased relative delay in starting tasks
across larger fractions of allocated resources.

There are systems that implement a specific set of the Pilot abstraction's
functionality to fulfill specific needs. For example, Pegasus-MPI-Cluster
(PMC)~\cite{Rynge:2012jj}, which is an MPI-based Master/Worker framework that
can be used in combination with Pegasus. This enables Pegasus to run
large-scale workflows of small tasks on HPC resources, but the tasks are
limited to single-node execution. In addition, there is a dependency on
\(fork()\)/\(exec()\) on the compute node which certain HPC resources do not
support (e.g. IBM BG/Q).

The Pilot paradigm has proven sufficiently useful for certain types of
workloads that resource manager systems have begun to include pilot
capabilities either as separate tooling, or as part of their implementation.
For example, CRAM~\cite{cram-scicomp14} is a tool developed specifically to
execute static ensembles of MPI tasks on HPC resources. Developed for
Sequoia, an IBM BG/Q system at LLNL, CRAM parallelizes the execution of an
application with many input parameters by statically bundling it into a
single MPI executable.

Flux is described as a next-generation Scalable Resource and Job Management
Software (RJMS) for HPC centers that focuses on a new paradigm of resource
and job management. Within this new paradigm, Flux allows resource allocation
to be dynamic (i.e., dynamic workloads), a key design principle of the Pilot
paradigm. This results in jobs having the ability to scale up to a maximum
requested during execution. Unfortunately, Flux is limited only to those HPC
resources that use it as their RJMS\@. Further, as of the writing of this
paper, Flux is still on an Alpha release, not suitable for production use.

As illustrated above, there has been various implementations of Pilot systems
in the past, but such systems tend to be limited to certain resources,
specialized to specific types of workloads, or a combination of tools is
needed to implemented the Pilot abstraction's functionality. Consequently,
many of these systems are limited in their ability, if able at all, to
support the execution of workloads comprised of multi-core and multi-node
tasks of many minutes to hour-long duration on HPC machines. We address these
limitations via RADICAL-Pilot (RP), a Pilot system designed to natively and
effectively support these types of workloads at scale.

\section{Design of RADICAL-Pilot}\label{sec:arch}

RADICAL-Pilot (RP) is a runtime system designed to execute multiple types of
scientific workloads on pilots instantiated on one or more resources. RP
enables the description of generic workloads with one or more scalar, MPI,
OpenMP, multi-process, and multi-threading tasks. These tasks can be executed
on CPUs, GPUs and other accelerators, on the same pilot or across multiple
pilots. We focus the discussion of design and architecture for HPC platforms.

RP implements two main abstractions: Pilot and Compute Unit (CU). Pilots and
CUs abstract away specificities of resources and workloads, making it
possible to schedule workloads either concurrently or sequentially on
resource placeholders. Pilots are placeholders for computing resources,
where resources are represented independent from architecture and topological
details. CUs are units of work (i.e., tasks), specified as an application
executable alongside its resource and execution environment requirements.

As a runtime system, RP offers an API to describe both pilots and CUs,
alongside classes and methods to manage acquisition of resources, scheduling
of CUs on those resources, and the staging of input and output
files. Reporting capabilities update the user about ongoing executions and
profiling capabilities enable detailed postmortem analysis of workload
executions and runtime behavior.

\subsection{Architecture and Implementation}

\begin{wrapfigure}{R}{0.5\textwidth}
\UPP{}
  \centering
  \includegraphics[trim=0 0 0 0,clip,width=0.49\textwidth]{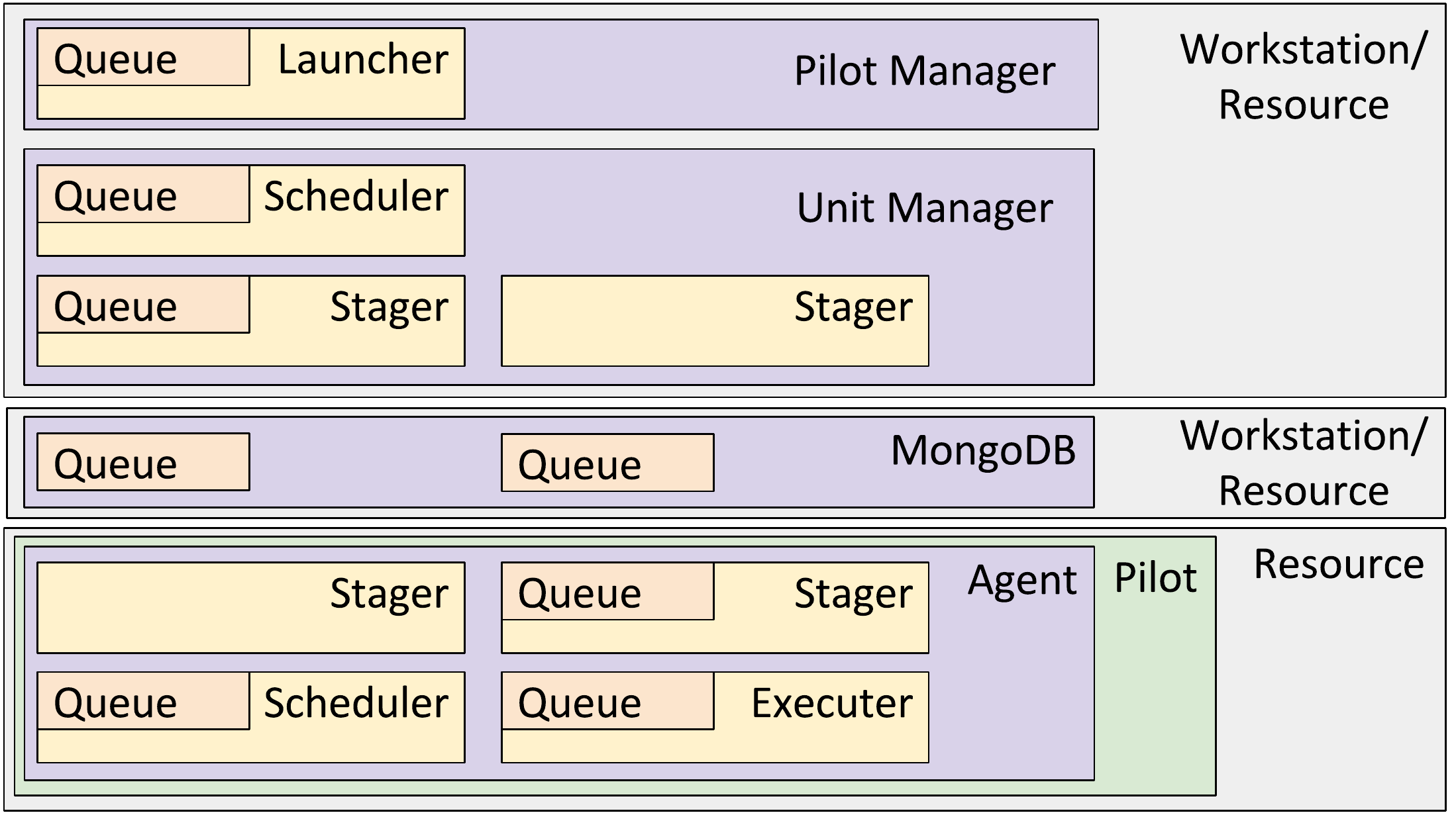}
  \caption{RADICAL-Pilot architecture.}\label{fig:arch-overview}
\UPP{}
\end{wrapfigure}

RP is a distributed system with four modules: PilotManager, UnitManager,
Agent and DB (Fig.~\ref{fig:arch-overview}, purple boxes). Modules can
execute locally or remotely, communicating and coordinating over TCP/IP\@,
and enabling multiple deployment scenarios. For example, users can run
PilotManager and UnitManager locally, and distribute DB and one or more
instances of Agent on remote computing infrastructures. Alternatively, users
can run all RP components on a remote resource.

PilotManager, UnitManager and Agent have multiple components, isolated into
separate processes. Some of the module's components are used only in specific
deployment scenarios, depending on both workload requirements and resource
capabilities. Components are stateless and some of them can be instantiated
concurrently to enable RP to manage multiple pilots and units at the same
time. This enables scaling of throughput and tolerance to failing components.
Concurrent components are coordinated via a dedicated communication mesh,
which introduces runtime and infrastructure-specific overheads, but improves
overall scalability of the system and lowers component complexity. Components
can have different implementations, configuration files can tailor RP toward
specific resources types, workloads, or scaling requirements.

PilotManager has a main component called `Launcher'
(Fig.~\ref{fig:arch-overview}). The Launcher uses resource configuration
files to define the number, placement, and properties of the Agent's
components of each Pilot. Currently, configuration files are made available
for essentially all US NSF and DOE production resources as well as Beowulf
variants, but users can provide new files or alter existing configuration
parameters at runtime, both for a single pilot or a whole RP session.

Agent has four main components: one Stager for input and output data,
Scheduler and Executor (Fig.~\ref{fig:arch-overview}).  Multiple instances of
the Stager and Executor components can coexist in a single Agent. Depending
on the architecture of the resource, the Agent's components can individually
be placed on cluster head nodes, MOM nodes, compute nodes, virtual machines,
or any combination thereof. ZeroMQ communication bridges connect the Agent
components, creating a network to support the transitions of the units
through components.

Once instantiated, each Agent's Scheduler gathers information from the
resource manager (RM) retrieving the number of CPUs (cores) and GPUs held by
the pilot on which the Agent is running and how those cores are partitioned
across nodes. Currently, the Scheduler acquires information from physical or
virtual Linux machines and the following RMs: TORQUE, PBS Pro, SLURM, SGE,
LSF, LoadLeveler, and Cray CCM\@.

Depending on requirements, the Agent's Scheduler assigns cores and GPUs from
one or more nodes to each unit.  For example, cores on a single node are
assigned to multithreaded units while, cores on topologically close nodes are
assigned to MPI units to minimize communication overheads. Two scheduling
algorithms are currently supported: ``Continuous'' for nodes organized as a
continuum, and ``Torus'' for nodes organized in an n-dimensional torus, as
found, for example, on IBM BG/Q.

The Agent's Scheduler passes the units on to one of the Agent's Executors,
which use resource configuration parameters to derive the launching command
of each unit. Currently, RP supports the following launching methods: MPIRUN,
MPIEXEC, APRUN, CCMRUN, RUNJOB, DPLACE, IBRUN, ORTE, RSH, SSH, POE, and
FORK\@. Among these, ORTE (Open RunTime Environment) enables scaling pilots
on leadership-class machines beyond the limited amount of concurrent process
allowed by methods like APRUN, MPIEXEC or MPIRUN\@.

Once the launching command is determined and further qualified depending on
the unit parameters and on the characteristics of the execution environment,
the Agent's Executors will execute those commands to spawn the application
processes.  Different spawning mechanisms are available: ``Popen'' (based on
Python), ``Shell'' (based on \T{/bin/sh}, and ``ORTELIB'' (based on the
\T{libopen-rte} API, bound to Python via CFFI).  Executors monitor the
execution of the units, collect exit codes, and communicate the freed cores
to the Agent's Scheduler.

\subsection{Execution Model}\label{sub:modcomp}

Workloads and pilots are described via the Pilot API and passed to the RP
runtime system (Fig.~\ref{fig:arch-overview}, 1). The PilotManager submits
pilots on one or more resources via the SAGA API
(Fig.~\ref{fig:arch-overview}, 2). The SAGA API implements an adapter for
each supported resource type, exposing uniform methods for job and data
management. Once a pilot becomes active on a resource, it bootstraps the
Agent module (Fig.~\ref{fig:arch-overview}, 3).

\begin{wrapfigure}{R}{0.5\textwidth}
\UPP{}
  \centering
  \includegraphics[trim=0 0 0 0,clip,width=0.49\textwidth]{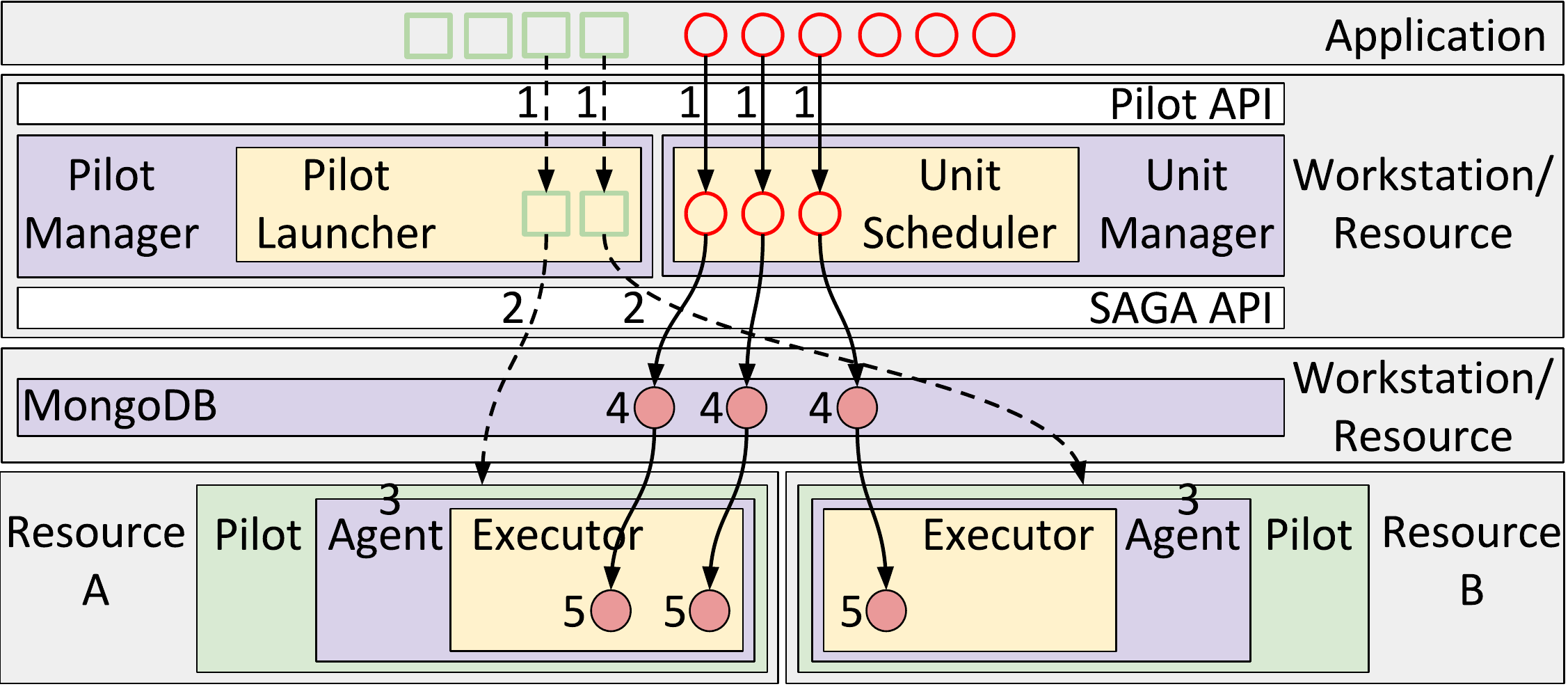}
  \caption{RADIAL-Pilot execution model.}\label{fig:exec-model}
\UPP{}
\end{wrapfigure}

The UnitManager schedules each unit to an Agent
(Fig.~\ref{fig:arch-overview}, 4) via a queue on a MongoDB instance. This
instance is used as the RP DB module to communicate the workload between
UnitManagers and Agents. Each Agent pulls its units from the DB module
(Fig.~\ref{fig:arch-overview}, 5) schedules each unit on the Executor when
the required amount of computing resources are available (e.g., number of
cores or GPUs). The Executor sets up the unit's execution environment and
then spawns the unit for execution.

Once a unit returns from its execution, the Executor communicates to the
Scheduler that resources have been freed. Scheduling loop can then proceed
until no units are left to execute. Once all the workload has been executed,
the runtime system is terminated to avoid inefficiencies in the use of
resource allocation. Multiple workloads can be executed within the time
boundaries during which resource are available, determined by constraints,
for example, walltime, preemption, or cost.

When required, the input data of a unit are either pushed to the Agent or
pulled from the Agent, depending on data locality and sharing requirements.
Similarly, the output data of the unit are staged out by the Agent and
UnitManager to a specified destination, e.g., a filesystem accessible by the
Agent or the user workstation. Both input and output staging are optional,
depending on the requirements of the units. The actual file transfers are
enacted via SAGA, and currently support (gsi)-scp, (gsi)-sftp, Globus Online,
and of course local filesystem operations.

\subsection{Profiling}\label{sub:profiling}

The distributed, modular, and concurrent design of RP introduces complexities
with both usability and performance overheads. The performance overheads of
RP's design require experimental characterization as they depend on the
properties of both the workloads and the resources used for the execution.
The execution overheads introduced at resource level are particularly
relevant as they affect the execution of every unit, independent of whether
the workload is divided in stages, or bounded by task dependences. The
overheads introduced by bootstrapping the components of the Agent, scheduling
the units, and launching them contribute to the overall time to completion of
the workload, and also affect overall resource utilization.

The characterization of RP performance requires dedicated capabilities. We
developed a profiler to enable postmortem performance analysis. The RP
profiler collects up to 200 unique events across all module components. Each
event contains a timestamp; an event identifier; component process and thread
IDs; IDs of pilots and units involved. Each time stamp is recorded
asynchronously to disk so as to minimize the overhead of the profiling. The
resulting profiles provide complete, timestamped traces for most operations
on any of the entities managed by RP\@: pilots, units and input/output files.

Profiling adds a certain overhead, and results as presented in this
publication all include that overhead.  By using buffered I/O and small data
structures we can keep the overhead manageable: a typical run from the
scaling experiments has pilot runtime of \(1045.5 \pm 29.4s\) without
profiling, and of \(1069.2 \pm 49.5s\) when profiling is turned on. Profiling
thus increases the runtime of about \(2.5\%\), and also slightly increases
the noise of the measurements.

A challenge in analyzing profiles from distributed systems such as RP is
clock synchronization.  Our post-mortem analysis toolchain called
RADICAL-Analytics (RA)~\cite{radical_analytics_url}) synchronizes profile
timestamps via NTP synchronization points.  It also performs consistency
checks on the profile data, and runs different time series analysis to
provide detailed insight into RPs runtime behavior.  We use RA in
Section~\ref{sec:exp} to characterize RP's performance.

\section{Performance Characterization}\label{sec:exp}

In this section, we focus on characterizing the performance of RP when
executing workloads requiring an ensemble of up to 16,000 independent MPI
tasks. Such workloads are representative of several scientific domains, and
pose unprecedented computational challenges due to the number of tasks, even
though the size of individual MPI tasks might be modest. The scalable
execution of such workloads requires leadership-class HPC machines, which are
typically designed and often optimized to support the execution of single,
very large MPI jobs, making it challenging to coordinate the concurrent and
sequential execution of thousands of modest sized MPI tasks.

The Pilot abstraction addresses these challenges by decoupling the
acquisition from the assignment of resources: once resources are acquired via
a single large pilot job, a pilot Agent is used to schedule and manage the
execution of tasks on those resources. We characterize the scaling and
performance of RP in terms of mean time to execution (TTX) of a workload and
the computing resource utilization (RU).

\subsection{Experiments Design}\label{ssec:exp_design}

The execution of workloads requires the interplay of all RP components and
their supporting infrastructure. Nonetheless, as seen in~\S\ref{sec:arch},
Figures~\ref{fig:arch-overview}-\ref{fig:exec-model}, RP reduces every
workload down to the execution of a set of compute units on an Agent. The
Agent retrieves units individually or in bulk and executes them on its
resources. As such, the characterization of TTX and RU depends on how each
Agent component performs.

As explained in~\S\ref{sec:intro} and~\S\ref{sec:arch}, the Pilot abstraction
and the RP Agent enable the execution of tasks both concurrently and
sequentially. Above a certain number of tasks, the workload cannot be
executed with full concurrency, even on the largest HPC machines currently
available, In this situation, sequential ``batched'' execution incur
overheads determined by the systems and resources used to manage the
execution.

Our experiments are designed to measure the overhead that the Agent,
third-party libraries, and the resources add to the execution of the
workload. Overhead captures the time spent either waiting for the workload's
computation to start, or performing operations other than those required by
the workload.  This overhead determines a partial utilization of the
available computing time for executing the workload and, therefore, a certain
degree of inefficiency of its execution.  We investigate its growth with
increasing number of units and cores.

We designed two experiments to measure the overhead of the Agent when
executing the workloads described in~\S\ref{sec:intro}. The first experiment
measures the weak scaling properties of the Agent by maintaining a constant
ratio of units to resources. The second experiment measures the strong
scaling of the Agent measured by fixing the number of units while varying
resources.

\begin{wrapfigure}{R}{0.5\textwidth}
\UPP{}
  \centering
  \includegraphics[trim=0 0 0 0,clip,width=0.49\textwidth]{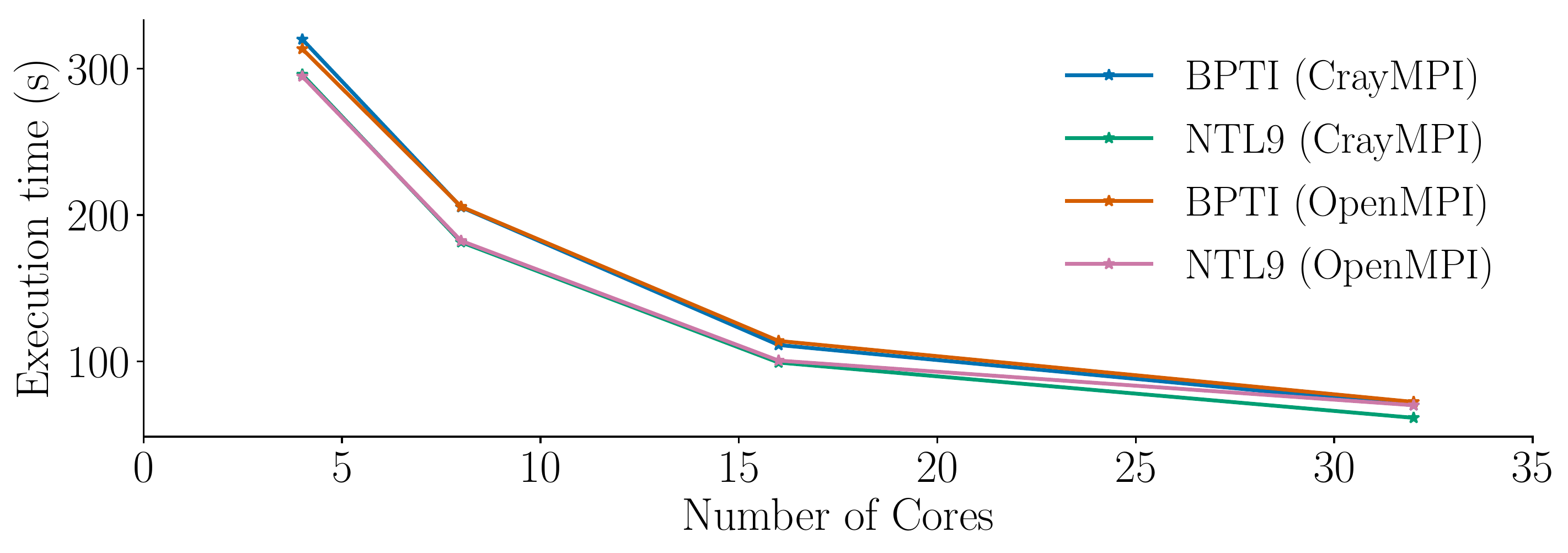}
\up{}
  \caption{BPTI, NTL9 scaling on Titan.}\label{fig:bpti-scaling}
\UP{}
\end{wrapfigure}

The experiments use a workload comprised of tasks involving the MD
simulations of the bovine pancreatic trypsin inhibitor (BPTI), a globular
protein of 20521 atoms when fully solvated. Figure~\ref{fig:bpti-scaling}
shows the scaling behavior of GROMACS for exemplar workloads and its
suitability for multi-node executions on Titan: although the simulations of
BPTI and NTL9 (14100 atoms when fully solvated) scale sublinearly after 8
cores, 32 cores offer the best relative performance, as measured by execution
time.

MD simulations with multiple GROMACS tasks executed on HPC machines can
experience large performance fluctuations over the mean runtime values. Such
fluctuations would make the separation of RP overheads from resource
fluctuations and runtime variations of the application's tasks difficult, if
not impossible. Thus, we profiled and emulated GROMACS simulations with
Synapse~\cite{merzky2016synapse}. Synapse profiles the compute, memory and
I/O use of an executable and emulates them. It reproduces the computing
activities of an executable, faithfully approximating its time to completion
and resource utilization.

Synapse offers our experiments several advantages over the direct use of the
executable it emulates: (1) simplified and self-contained deployment without
third parties libraries and compilers dependences; (2) high-fidelity
replication of the computing patterns of the emulated executable without
actual input/output files; (3) profiling capabilities independent of third
parties applications; (4) control over the number of FLOPs executed; and (5)
selective emulation of the type of profiled resources. As such, Synapse
allows greater control, while simplifying deployment and data analysis
without loss of generality of results.

\begin{wrapfigure}{R}{0.5\textwidth}
\UPP\up{}
  \centering
  \includegraphics[trim=0 0 0 0,clip,width=0.49\textwidth]{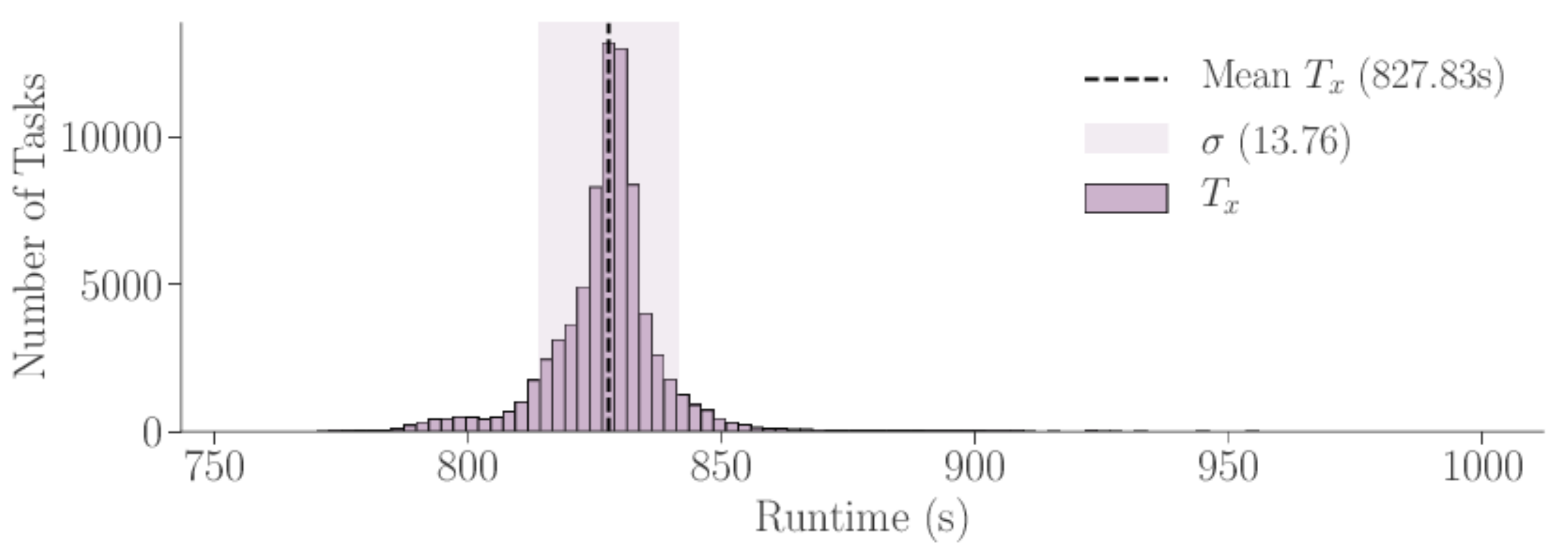}
\UP{}
  \caption{Distribution of the TTX for Synapse emulation of BPTI.}\label{fig:tx_distribution}
\UPP{}
\end{wrapfigure}

We emulated the execution of a single GROMACS instance, simulating BPTI for
\(\approx\)250ps, the baseline in several studies. In this way, we controlled
the runtime noise inherent to executing multiple instances of the same
executable: we measured only the variance of Titan and the predictable
variance of Synapse. Further, we did not emulate I/O activities as the
performance fluctuations of Titan's network file systems would have dominated
our experimental results. Figure~\ref{fig:tx_distribution} shows the narrow
distribution of Synapse emulations' runtime: the mean is 828s with a standard
deviation of \(\pm\)14s.

\begin{table*}[!b]
	\caption{Experiments 1 and 2: Weak and strong scalability.}\label{tab:experiments}
	\centering
	\resizebox{\textwidth}{!}{%
	\begin{tabular}{p{0.7cm}       
	                p{2.5cm}       
	                p{2.3cm}       
	                p{1.8cm}       
	                p{1.7cm}       
	                p{2.5cm}       
					}
	\toprule
    \B{ID}                      &  
    \B{\#Tasks}                 &  
    \B{\#Gener-\newline ations} &  
    \B{Task \newline Runtime}   &  
    \B{\#Cores/ \newline Task}  &  
    \B{\#Cores/ \newline Pilot} \\ 
	\midrule
	\B{1}                      &   
	$2^n, n=[5-12]$            &   
	1                          &   
	\mr{828s\(\pm\)14s}        &   
	\mr{32}                    &   
	\(2^n, n=[10-17]\)         \\  
	%
	%
	\B{2}                      &   
	\(2^{14}\)                 &   
	\(2^n, n=[5-3]\)           &   
	&
	&
	\(2^n, n=[14-16]\)         \\  
	\bottomrule
	\end{tabular}
	}
\end{table*}

Table~\ref{tab:experiments} shows the 8 runs of Experiment 1, designed to
measure weak scaling of RP with the chosen workload on Titan. Each run
executes between 32 and 4,096 32-cores tasks on a single pilot with between
1,024 and 131,072 cores. The ratio between the number of tasks executed and
the amount of resources acquired is constant across the 8 runs of the
experiment. All the tasks are thus executed concurrently in a single
so-called `generation', i.e., a single set of concurrent executions. As all
the tasks have analogous overheads and all the tasks execute concurrently,
the median of the ideal total execution time (TTX) of all the tasks should be
analogous for all the 8 runs.

Table~\ref{tab:experiments} shows the 3 runs for Experiment 2, designed to
measure the strong scaling of RP with the chosen workload on Titan. Different
from Experiment 1, the ratio between number of tasks and number of cores of
the pilot varies: Each run executes 16,384 tasks on a single pilot with
between 16,384 and 65,536 cores. The rest of the parameters are the same as
Experiment 1, each task with a mean execution time of 828s\(\pm\)14 and
requiring 32 cores. Because of the disparity between the number of cores
required by the tasks and the number of cores of the pilot, the workload is
executed on multiple generations, between 32 and 8.

\subsection{Weak and Strong Scaling}\label{ssec:exp_ws_ss}

\begin{figure}
  \centering
  \UP\up
  \includegraphics[trim=0 0 0 80,clip,width=0.49\textwidth,valign=t]{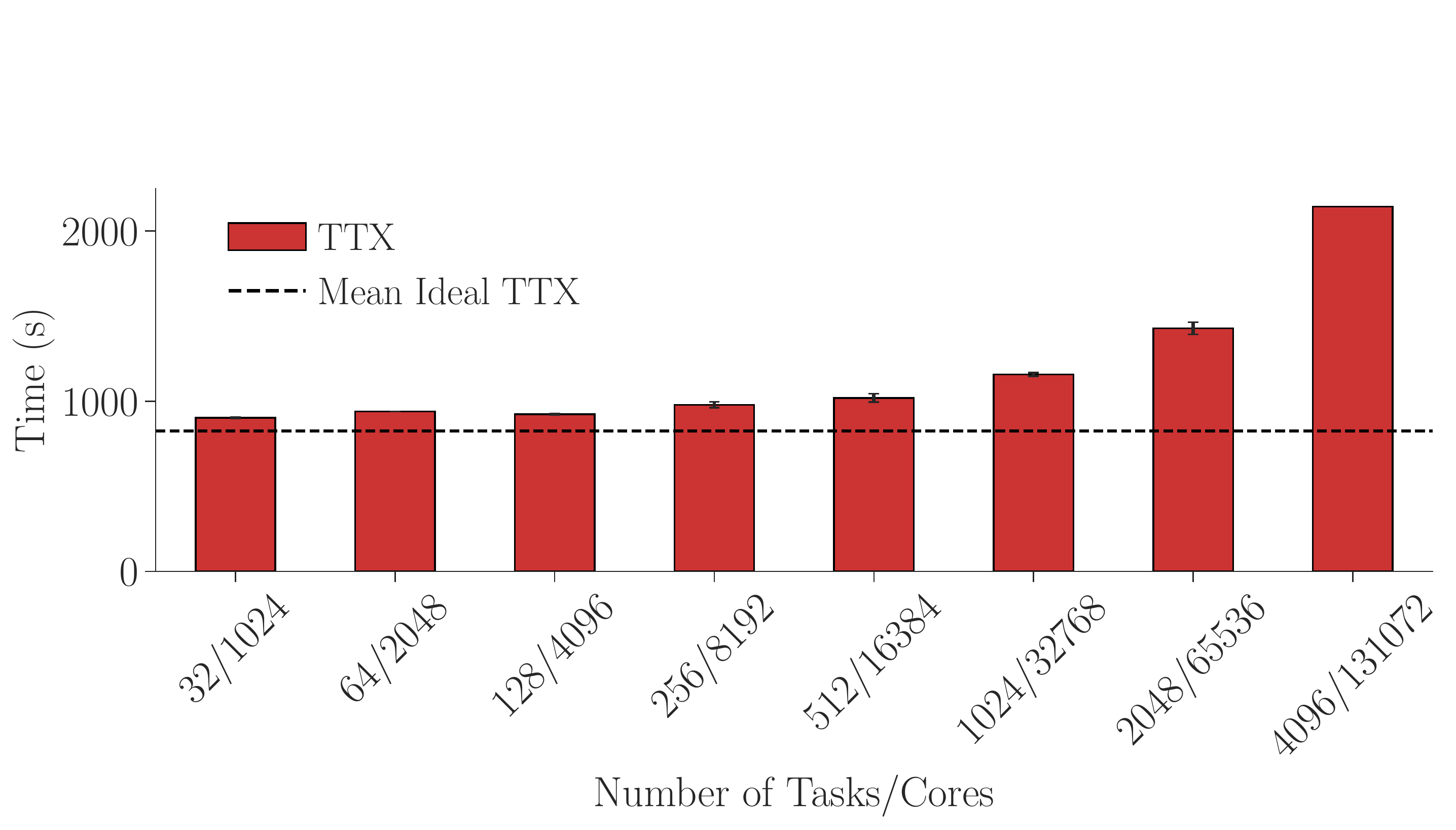}
  \includegraphics[width=0.49\textwidth,valign=t]{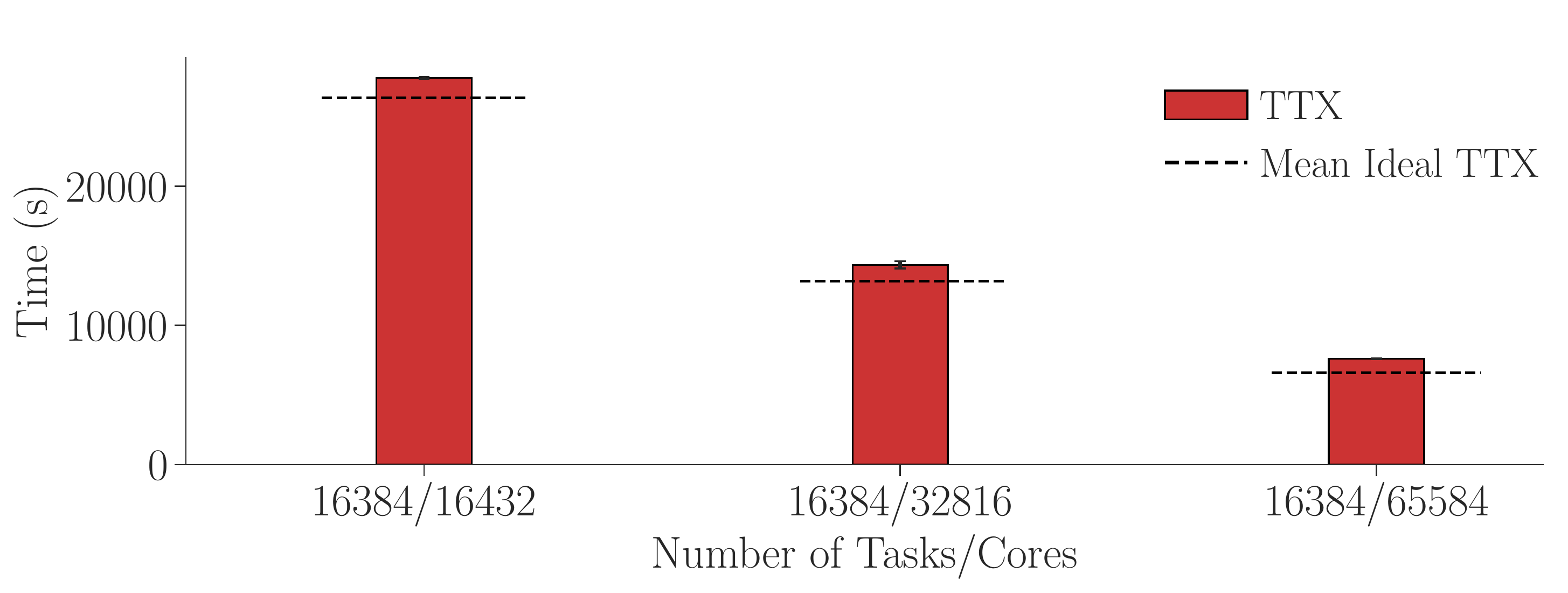}
  \UP
  \caption{\textbf{Experiments 1 and 2:} Weak (left) and Strong (right)
  scaling of RADICAL-Pilot.}\label{fig:s-ttc}
  \UPP
\end{figure}

Figure~\ref{fig:s-ttc}~(left) shows the weak scaling of RP for the workload
described in Table~\ref{tab:experiments}. An ideal TTX (broken line)
represents execution time without RP and resource overheads and corresponds
to the mean value in Figure~\ref{fig:tx_distribution}. In Experiment 1
(Table~\ref{tab:experiments}), the ratio between number of tasks and core is
constant, enabling fully concurrent executions.

Figure~\ref{fig:s-ttc}~(left) shows that the actual TTX plotted on the right
axes, scales almost linearly between 1,024 and 4,096 cores, and sublinearly
between 4,096 and 131,072 cores. The average value of TTX for runs with
between 1,024 and 4,096 cores is 922s\(\pm\)14, indicating an average
overhead of 11\% over the mean of the ideal TTX\@. This overhead grows to
18\% at 8,192, and 160\% at 131,072 cores.

Figure~\ref{fig:s-ttc}~{(right)} shows that the strong scaling of 16,384
tasks executed from 16,384 to 65,536 cores; this results in the number of
generations varying from 32 to 8. When executed over 16,384, 32,816 and
65,536 cores, they have a TTX of 27,794s\(\pm\)70, 14,358s\(\pm\)259, and
7,612s\(\pm\)29 respectively. The deviation from ideal TTX is relatively
uniform across different pilot sizes---1,158s\(\pm\)150, which indicates that
RP is relatively less efficient at higher pilot core counts.

\begin{wrapfigure}{R}{0.5\textwidth}
\UPP\up
  \centering
  \includegraphics[trim=0 0 0 25,width=0.49\textwidth]{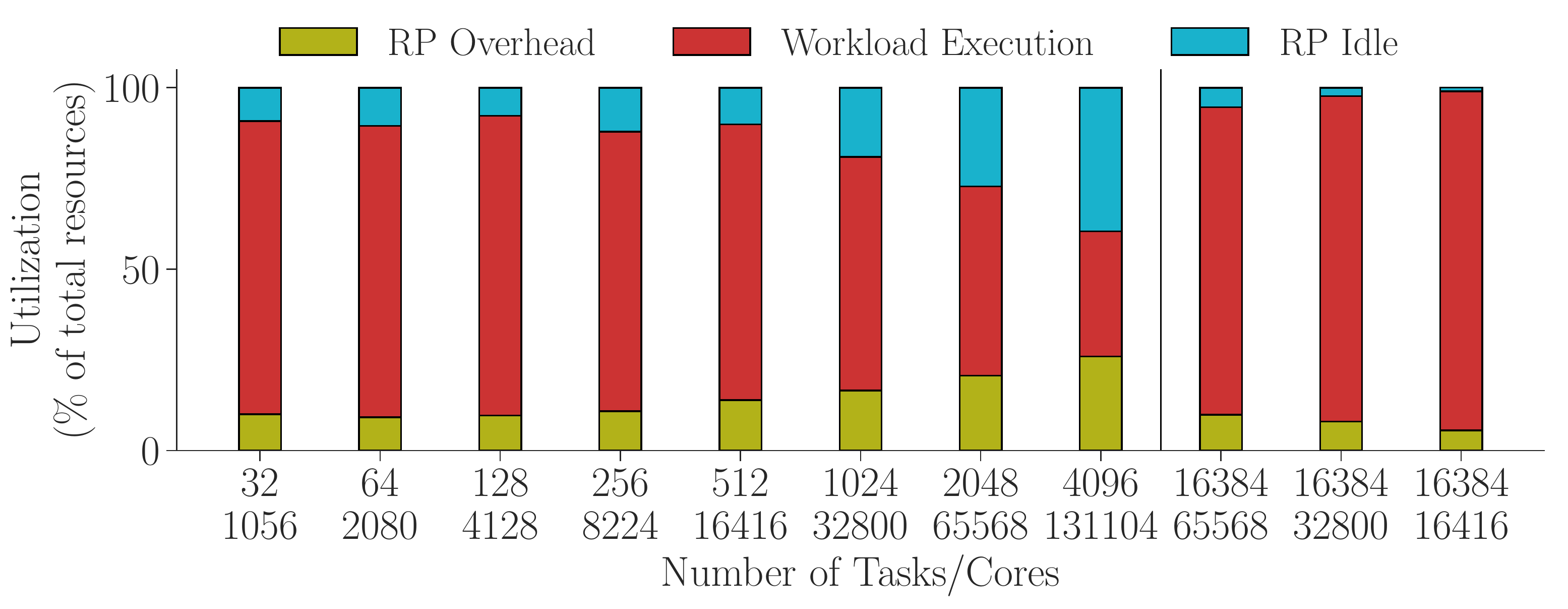}
\UP{}
  \caption{Experiment 1 and 2: Resource utilization of RADICAL-Pilot.}\label{fig:ws-ru}
\UPP
\end{wrapfigure}

Figures~\ref{fig:ws-ru} shows the resource utilization (RU) in terms of the
percentage of the available core-time spent executing the workload (red), RP
code (green), or idling (blue) for Experiment 1 (first 8 bars) and Experiment
2 (last 3 bars). Note the relation between TTX and RU\@: The more core-time
is spent executing the workload, the shorter TTX\@. Figure~\ref{fig:ws-ru}
shows a relatively constant percentage of core-time utilization for runs with
between 32/1024 and 128/4096 tasks/cores, consistent with TTX of
Figure~\ref{fig:s-ttc}~(left). The percentage of utilization decreases with
the growing of the number of tasks/cores, also consistent with
Figure~\ref{fig:s-ttc}~(left).

The last three bars of Figure~\ref{fig:ws-ru} show progressively shorter
values for both RP overhead and idling for runs with multiple generations (as
defined in \S\ref{ssec:exp_design}). When tasks of one generation terminate,
those of the following generation immediately starts executing. This
eliminates the idling of cores for all generations but the last one. Further,
RP overhead increases with the number of cores, indicating that the reduced
performance of RP measured in Figure~\ref{fig:s-ttc}~(left) depends, at least
to some extent, on the size of the pilot used.

\subsection{Understanding Weak Scaling}\label{ssec:exp_agent}

As seen in~\S\ref{sec:arch}, the RP Agent is implemented with multiple
components that can execute concurrently, depending on the workload and the
resources on which the Agent is running. Further, multiple instances of the
same component can be created to concurrently manage subsets of available
tasks and resources. Although distribution and concurrency improve
performance, they also make it difficult to determine the underlying causes
of sublinear weak scaling.

\subsubsection{Component Concurrency}

\begin{figure*}
\UPP{}
  \centering
  \includegraphics[trim=0 0 0 80,clip,width=\textwidth]{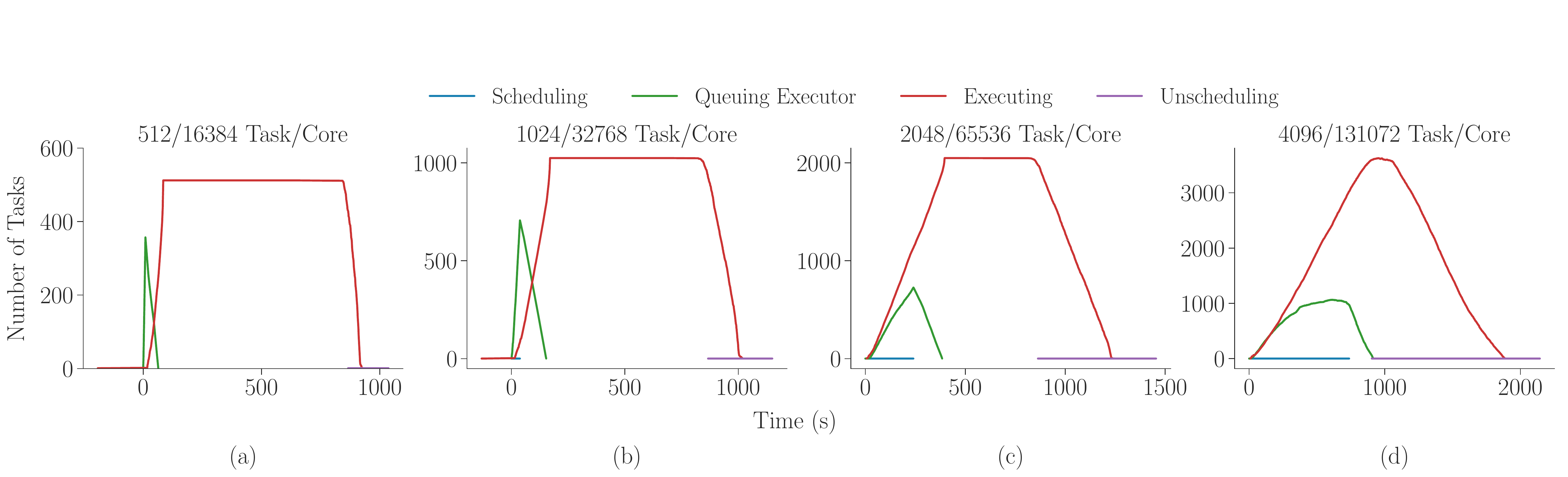}
\UP{}\UP{}
  \caption{Task concurrency weak scaling: Scheduler and Executor components.}\label{fig:ws-concurrency}
\UPP{}
\end{figure*}

Figure~\ref{fig:ws-concurrency} shows the concurrency of the two main
components of the Agent (Scheduler and Executor) and their queues.
Concurrency is expressed as the number of tasks managed by each component (or
their queue) at any point in time. Each task is initially handled by the
Scheduler (Scheduling, Blue); as soon as the required amount of resources for
the task become available, it is queued to the Executor (Queuing Executor,
Green). Once in the Executor, the task is executed (Executing, Red). Once
the execution completes, the task is marked as Done, Cancel or Failed and the
Scheduler is informed that resources have been freed (Unscheduling, Purple).

The Scheduling and Unscheduling operations have concurrency value of 1
because the Scheduler handles one task at a time, and does not retain any
tasks in the single-generation setup used for the weak scaling experiment.
The time taken to schedule and unschedule tasks (lengths of the blue and
purple lines) increases with scale. The number of tasks concurrently queued
in the Executor decreases with scale (height of the green line), while the
time spent in the queue by the tasks increases (X-value of end of the green
line). Similarly, the duration of the plateau (i.e., the maximum number of
tasks handled by the Executor as represented by the red line) gets
progressively smaller with scale. Figure~\ref{fig:ws-concurrency}(d) shows
that at 4096/131072 cores/tasks, RP does not reach maximum concurrency: the
height of red line does not reach 4096 concurrent tasks.

\subsubsection{Scheduler and Executor Overheads}\label{sssub:overhead}

\begin{figure*}
\UPP{}
  \centering
  \includegraphics[trim=0 0 0 80,clip,width=\textwidth]{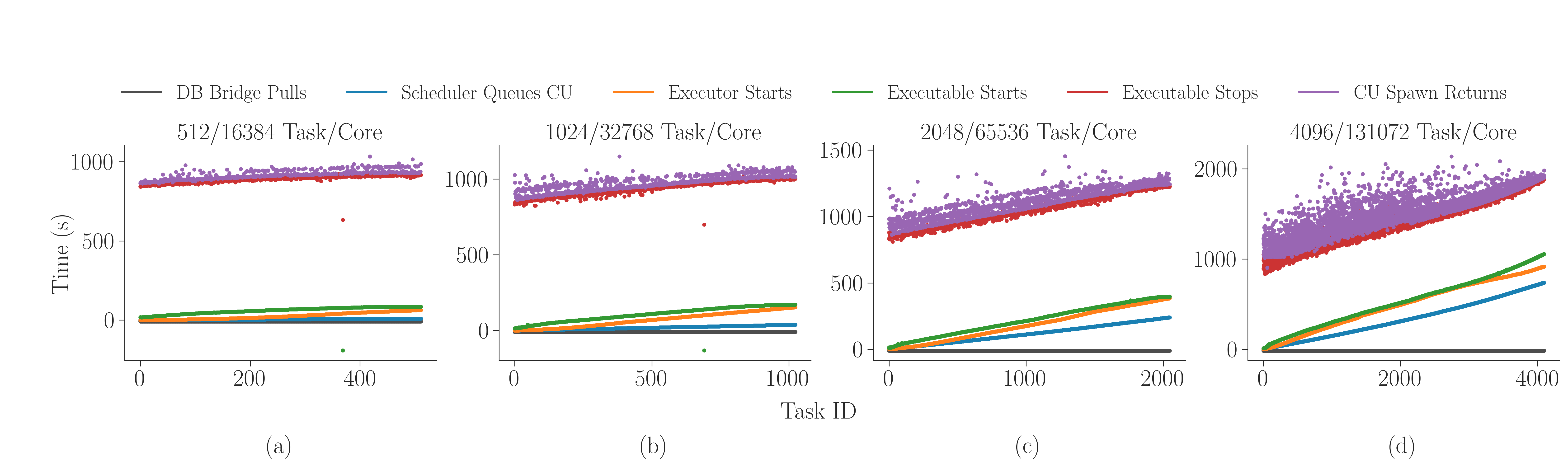}
\UP{}\UP{}
  \caption{Task events weak scaling: Scheduler and Executor
  components.}\label{fig:ws-events}
\UPP{}
\end{figure*}

Figure~\ref{fig:ws-events} helps clarify the relation between the performance
of the Scheduler and Executor, the two Agent components that appear to
contribute to RP overhead. We measure the time spent by each task in each
component of the RP Agent. Tasks are pulled from RP DB into Scheduler's queue
(DB Bridge Pulls); after scheduling, the Scheduler queues each task into
Executor (Scheduler Queues CU); Executor starts processing the queued task
(Executor Starts); the task's executable starts (Executable Starts) and stops
(Executable Stops) executing; and, finally, Executor marks the task as done
(CU Spawn Returns).

Figures~\ref{fig:ws-events}(a)-(d) shows that all the tasks of the workload,
pulled in bulk from the DB (DB Bridge Pulls), enter Scheduler's queue
approximately at the same time (i.e., all the tasks are approximately at the
same height compared to the y-axes, forming an almost horizontal line,
parallel to the x-axis). The Scheduler takes comparatively more time to
schedule the tasks in Figures~\ref{fig:ws-events}(d) than in
Figures~\ref{fig:ws-events}(a)-(c). Note how the slope of the line formed by
the events becomes increasingly steeper from Figures~\ref{fig:ws-events}(a)
to Figures~\ref{fig:ws-events}(d). The mean scheduling time for 512 tasks on
16,384 cores is 18s; 39s with 1,024/32,768 tasks/cores; 129s with
2,048/65,536 tasks/cores; and 350s with 4,096/131,072 tasks/cores.

Figures~\ref{fig:ws-events}(a)-(d) also show two overheads in the Executor:
(1) the time spent to prepare a task for its execution (Executor Starts),
i.e., the time between when a task is passed to ORTE and when it starts to
execute; and (2) the time required for the Executor to be informed that a
task has been executed (CU Spawns Return), i.e., the time from when a task
stops executing and the time when ORTE passes a message to the Executor about
the task Done or Failed state. The mean time to prepare the execution of 512
tasks on 16,384 cores is 37s\(\pm\)9; 37s\(\pm\)6 with 1,024/32,768
tasks/cores; 35s\(\pm\)8 with 2048/65536 cores; and 41s\(\pm\)30 with
4,096/131,072 tasks/cores, which in spite of the high jitter, makes the mean
essentially invariant across scales.

The Executor takes variable amount of time to acknowledge that the execution
of a task has completed (as given by the timestamp of the CU Spawn Returns
event). This variance increases with scale, depending mostly on the time
taken by ORTE (i.e., the lunch method used on Titan to execute tasks) to
communicate with RP about the task's state. The distribution of the CU Spawn
Returns event is both broad and long-tailed across all the scales. The mean
time to communicate the completion of 512 tasks on 16,384 cores is
29s\(\pm\)16; 34s\(\pm\)28 with 1,024/32,768 tasks/cores; 59s\(\pm\)46 with
2048/65536 cores; and 135s\(\pm\)107 with 4,096/131,072 tasks/cores.

\subsubsection{Improving Scheduler Performance}

\begin{figure*}
\UPP{}
  \centering
  \includegraphics[trim=0 0 0 80,clip,width=\textwidth]{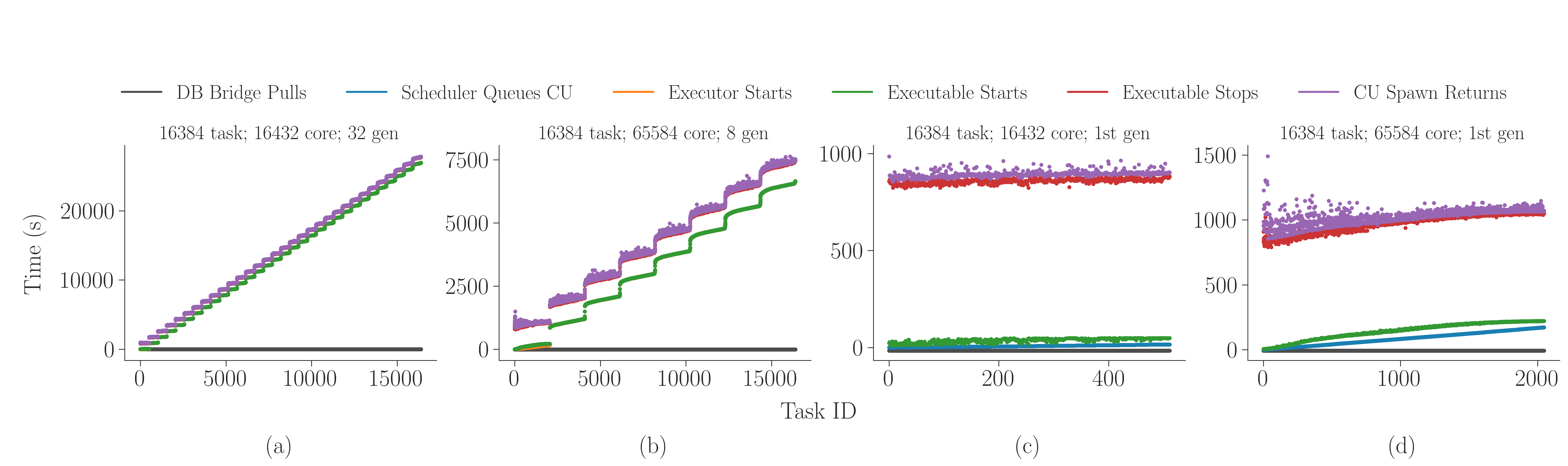}
\UPP{}
  \caption{Task events strong scaling: Scheduler and Executor
  components. Note: Scheduler Queues CU (blue) overlays Executor Starts
  (orange) events.}\label{fig:ss-events-1stgen}
\UPP{}
\end{figure*}

Figure~\ref{fig:ss-events-1stgen} confirms that Scheduler and Executor
performance mostly depend on the number of cores of the pilot.
Figure~\ref{fig:ss-events-1stgen}(a)-(b) show the strong scaling behavior of
RP with 32 and 8 generations (corresponding to 32 and 8 ``steps'' in (a) and
(b) respectively), as described in Table~\ref{tab:experiments}, Experiment 2.
Figure~\ref{fig:ss-events-1stgen}(c)-(d) plots the events of one of the 32
generations and those of one of the 8 generations respectively. The Scheduler
performance of the strong scaling runs measured: 23s\(\pm\)5 to schedule 512
tasks on 16,384 cores, 28s\(\pm\)13 2048 tasks on 32,768 cores, and
92s\(\pm\)49 2,048 tasks on 65,584 cores. This performance degradation
depends on the number of cores, as the number of tasks is constant across
Experiment 2 runs. The rate of degradation is similar to what we observed in
the weak scaling runs of comparable pilot size.

Scheduling overhead increases with the number of tasks scheduled: later tasks
have to wait longer to be handled, as the Scheduler works \I{sequentially}.
As a consequence, the performance of the default scheduler of RP decreases
with pilot size. This scheduler uses a scheduling approach where a Python
data structure representing the resource configuration is repeatedly searched
for free cores. This approach is needed and effective for a general purpose
scheduler to handle many different kinds of workloads: For example,
homogeneous/heterogeneous; MPI/OpenMP/Scalar; and single-node/multi-node.
However, as seen in~\S\ref{sssub:overhead}, the overhead of this approach
begin to dominate with more than 128 32-cores tasks and 4096 cores.

\begin{wrapfigure}{R}{0.5\textwidth}
\UP\up
  \centering
  \includegraphics[trim=0 0 0 25,clip,width=0.49\textwidth]{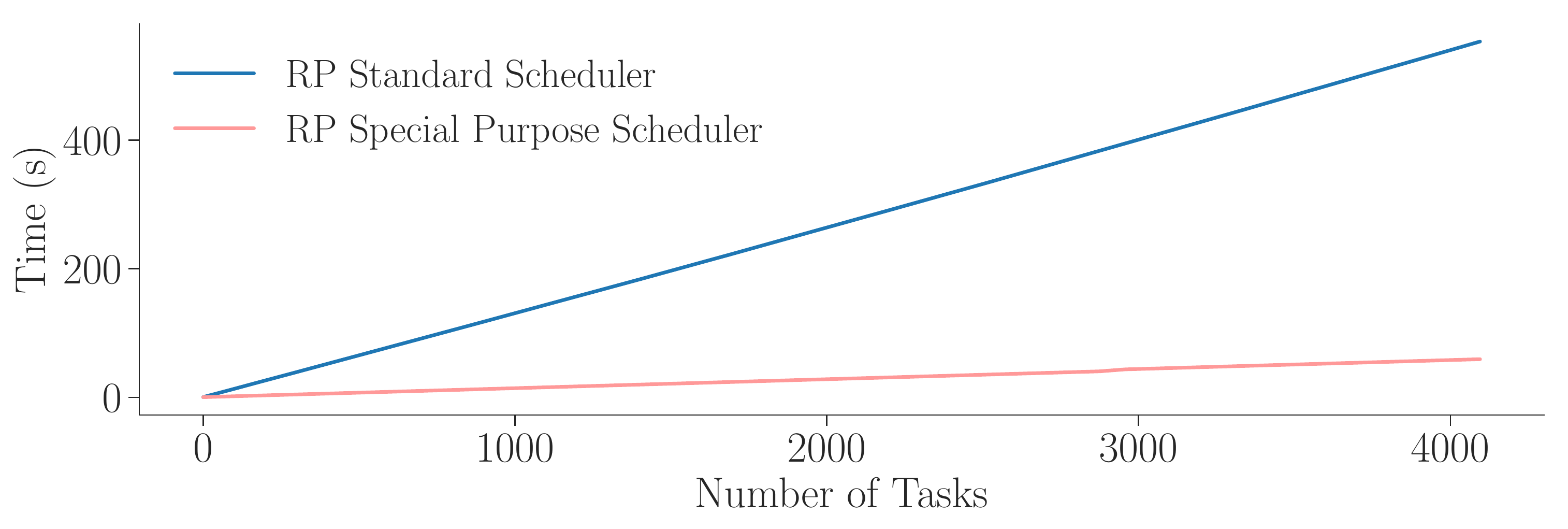}
\UP{}
  \caption{Scheduling overheads: Standard and special purpose
  schedulers.}\label{fig:scheduling_overheads}
\UP\up
\end{wrapfigure}

To address this, but also to demonstrate the flexibility and extensibility of
RP, we implemented a scheduler algorithm which handles the specific workload
used in our scaling experiments (homogeneous, multi-node MPI tasks). The
implementation of the special purpose scheduler takes as little as 30
lines of code.  Its behavior is shown in
Figure~\ref{fig:scheduling_overheads}: The scheduler now treats each task in
constant time, at a much lower time per task compared to the original
scheduler.

We reiterate that this performance improvement comes at the expense of
generality, i.e., the scheduler is constrained to handle only homogeneous
bag-of-tasks on homogeneous clusters. The loss of generality arises from a
change in critical code path from a search to a lookup, which improves the
scheduler throughput from 7 tasks/s to 70 tasks/s, an approximately 9 fold
improvement over the general purpose scheduler.

\section{Discussion and Conclusion}\label{sec:conclusion}

{\it Prima facie}, a software system implementing the Pilot
abstraction~\cite{review_pilotreview} provides the conceptual and functional
capabilities needed to serve as the runtime system for the scalable execution
of a workload comprised of many tasks. The impact of an abstraction is
limited to its best implementation. Whereas there are multiple Pilot systems,
they are geared towards either specific workloads or platforms. Against this
backdrop, RADICAL-Pilot (RP) brings together recent conceptual
advances~\cite{review_pilotreview} with systems \& software
engineering~\cite{bb_2016}.

This paper describes RP's design and implementation (Sec.~\ref{sec:arch}),
and characterizes the performance of its Agent module on Titan at the Oak
Ridge Leadership Computing Facility (Sec.~\ref{sec:exp}) for workloads
comprised of large number of modestly sized MPI tasks. Although RP works on
multiple platforms, we focus our experiments on Titan as it currently offers
the highest degree of concurrency (300K CPU cores) to researchers in the USA.
The experiments discussed in Sec.~\ref{sec:exp} benefited from RP's support
for introspection and profiling.  Using the RADICAL-Analytics toolchain to
analyze those profiles, we were able to pinpoint the main contributions to
RP's runtime overhead at scale (more than 40\% of Titan).

We demonstrated the extensibility of RP by implementing a simple dedicated
scheduler to address a very specific performance bottleneck: the default
internal RP scheduler was initially designed for generality, which came at a
price in performance. When constraints of generality were removed and the
internal scheduler was optimized for homogeneous set of MPI tasks, the
performance of the scheduler was significantly enhanced.

When using the optimized scheduler, both the performance and scalability of
RP are determined by system software limits. Specifically, we show that the
launch rate of tasks is dominated by an overhead arising from the use of
OpenMPI Runtime Environment (ORTE) of OpenMPI\@. Further, we observe that
failure rates in the ORTE layer increase significantly when utilizing 131K
cores and above. RP stresses ORTE capabilities, however the exact reasons for
high jitter in launch rates, and failures after 131K core are currently
unknown. Although, current capabilities support production requirements, we
continue to work with the ORTE developers to address the observed scalability
and stability issues.

The focus of this paper has been on the direct execution of workloads on HPC
machines, but RP also forms the middleware and runtime system for a range of
other tools and libraries~\cite{entk,repex2016,power-of-many17,extasy},
already used in production. RP is available for immediate use on many
contemporary platforms~\cite{radical_pilot_url}, 
accompanied with extensive documentation and an active developer-user
community.

For molecular sciences, there is an existing and demonstrated
need~\cite{nextgen-molsim} to support up to 10\(^5\)--10\(^6\) MPI tasks as
part of a single workload. The scalable execution of workloads comprised of
many heterogeneous tasks~\cite{mycray-2012} is an increasingly critical
requirement (see Ref.~\cite{LR-hpdc-2015} for a recent analysis on NERSC
systems). These workloads will range from a heterogeneous mix of tasks (both
spatial and temporal) to dynamically evolving workloads. RP's modular
architecture and extensible implementation enable us to balance generality
and performance, while promoting integration with application tools and
system software. As such, we consider RP and its software ecosystem to be
well-equipped to support these and future workloads on a wide variety of
platforms, including Summit at OLCF\@.

\vspace{-0.05in}
\subsection*{Software and Data} {\footnotesize Source code, 
raw data and analysis scripts to reproduce experiments can be found at:\\
RADICAL-Pilot: \url{https://github.com/radical-cybertools/radical.pilot}\\
RADICAL-Analytics: \url{https://github.com/radical-cybertools/radical.analytics}\\
Experiment data and scripts:
\url{https://github.com/radical-experiments/rp.paper}}

\vspace{-0.05in}
\subsection*{Acknowledgements} {\footnotesize This work is supported by
NSF ``CAREER'' ACI-1253644, NSF ACI-1440677 “RADICAL-Cybertools” and DOE
Award DE-SC0008651. This research used resources (Titan) at the Oak Ridge
Leadership Computing Facility at the Oak Ridge National Laboratory, which is
supported by the Office of Science of the U.S. Department of Energy under
Contract No. DE-AC05-00OR22725. We also acknowledge access to computational
facilities: XSEDE resources (TG-MCB090174), and Blue Waters (NSF-1713749).
We thank Mark Santcroos for his contributions to the RP-ORTE integration.}

\bibliographystyle{splncs}
\bibliography{rp}

\end{document}